\documentclass[prd,reprint,superscriptaddress,preprintnumbers,nofootinbib,amsmath,amssymb,aps]{revtex4-2}

\usepackage{ads}
\usepackage{graphicx}
\usepackage{dcolumn}
\usepackage{color}
\usepackage[colorlinks=true,citecolor=blue,urlcolor=blue]{hyperref}
\usepackage{import}

\begin{document}

\title{Reconstructing the early-universe expansion and thermal history}
\author{Rui An}
\thanks{Email: anrui@usc.edu}
\affiliation{Department of Physics and Astronomy, University of Southern California, Los Angeles, CA 90089, USA}
\author{Vera Gluscevic}
\thanks{Email: vera.gluscevic@usc.edu}
\affiliation{Department of Physics and Astronomy, University of Southern California, Los Angeles, CA 90089, USA}
\affiliation{TAPIR, Mailcode 350-17, California Institute of Technology, Pasadena, CA 91125, USA}

%%%%%%%%%%%%%%%%%%%%%%%%
\begin{abstract}
We present a model-independent reconstruction of the early expansion and thermal histories of the universe, obtained from light element abundance measurements. The expansion history is tightly constrained around the onset of the Big Bang Nucleosynthesis (BBN). The temperature of photons is additionally constrained around the time of neutrino decoupling. Allowing for perturbations to the standard expansion rate, we find that the radiation energy density is constrained to within 15\% of its $\Lambda$CDM value, and only 1\% extra matter energy density is allowed around the epoch of BBN. We introduce a new and general analytic fitting formula for the temperature variation, which is flexible enough to reproduce the signal of large classes of beyond-CDM particle models that can alter the temperature through early-time energy injection. We present its constraints from BBN data and from the measurements of effective number of relativistic species and helium-4 abundance probed by the Cosmic Microwave Background radiation anisotropy. Our results provide clarity on the most fundamental properties of the early universe, reconstructed with minimal assumptions about the unknown physics that can occur at keV--MeV energy scales and can be mapped to broad classes of models of interest to cosmology. 

\end{abstract}

\maketitle

%%%%%%%%%%%%%%%%%%%%%%%%%%%%
\section{Introduction}

Big Bang Nucleosynthesis (BBN) is a sensitive probe of the universe in the energy range from keV to MeV \cite{1996RPPh...59.1493S,2009PhR...472....1I,2010ARNPS..60..539P}. The observations of the abundances of light nuclei produced during BBN strongly constrain new physics occurring at these energies \cite{2020PTEP.2020h3C01P}. For example, a new light particle in thermal equilibrium at BBN would affect the expansion history and the temperature of the radiation, altering primordial element abundances away from the predictions of the Standard Model (SM) of particle physics \cite{1996PhR...267..195J,2014PhRvD..89h3508N,2015PhRvD..91h3505N}. BBN observations thus can constrain light dark matter (DM) particles, new neutrino species, and other relativistic degrees of freedom in the early universe \cite{2012arXiv1208.0032S,2013JCAP...08..041B,2014MmSAI..85..175S,2019JCAP...02..007E,2022PhRvL.129b1302G,2012CoPhC.183.1822A,2015PhLB..751..246K,2020JCAP...12..048K,2023PhRvD.107b3525S}. However, most data analyses assume a specific model for these deviations, enabling stringent, but model-dependent constraints. In this work, we use measurements of the light element abundances created in the process of BBN, and pursue a model-independent reconstruction of the early expansion and thermal histories, avoiding assumptions regarding the specific particle content of the universe.

In the standard cosmological model, the CMB blackbody evolution as a function of redshift leads to $T_\mathrm{CMB}(z) = T_\mathrm{CMB,0}(1+z)$ after electron-positron annihilation, where $T_\mathrm{CMB,0}$ is the CMB temperature today in the local universe \cite{2009ApJ...707..916F}. A departure from this specific redshift dependence would challenge the standard cosmological model. The CMB temperature was previously measured at low redshifts, using two methods: one is based on multi-frequency Sunyaev-Zeldovich (SZ) observations of galaxy clusters \cite{2005A&A...441..435H,2002ApJ...580L.101B,2009ApJ...705.1122L,2015ApJ...808..128D,2021ApJ...922..136L}; the other one relies on spectroscopic studies of absorption lines in quasar spectra \cite{1986ApJ...308L..37M,1994Natur.371...43S,2000Natur.408..931S,2001ApJ...547L...1G,2002A&A...381L..64M,2005ApJ...633..649C,2008A&A...482L..39S,2010A&A...523A..80N,2011A&A...526L...7N,2013A&A...551A.109M,2016PhRvD..93d3521A,2020AstL...46..715K}. A model-independent measurement of the CMB temperature in the early universe (at high redshift)---the goal of this work---is similarly of considerable cosmological interest, as it may probe physics at entirely different energy scales. 

We start from the current measurement uncertainties of the light element abundances, including helium $Y_\mathrm{p}$, deuterium $Y_\mathrm{D} \equiv (D/H)\times 10^5$, helium-3 ${}^3\!He/H$ and lithium-7 ${}^7\!Li/H$, and adopt a non-parametric model of the expansion and thermal histories. We then employ a simple Fisher martix analysis to identify the redshift range to which these measurements are most sensitive. Next, in order to constrain deviations in the histories away from the SM values at $z\gtrsim 10^7$, we perform a Markov Chain Monte Carlo (MCMC) analysis of the light element abundance data from the recent observations reported in Ref.~\cite{2020PTEP.2020h3C01P}, where they recommended $Y_\mathrm{p}=0.245\pm 0.003$ based on recent measurements \cite{2019ApJ...876...98V,2015JCAP...07..011A,2014MNRAS.445..778I}, $Y_\mathrm{D}=2.547\pm 0.025$ which is a weighted mean of the 11 most precise measurements \cite{2018ApJ...855..102C,2018MNRAS.477.5536Z,2017MNRAS.468.3239R,2016MNRAS.458.2188B,2015MNRAS.447.2925R,2014ApJ...781...31C,2016ApJ...830..148C}, ${}^3\!He/H = (1.1\pm 0.2)\times 10^{-5}$ taken from Ref.~\cite{2002Natur.415...54B, 2018AJ....156..280B}, and ${}^7\!Li/H = (1.6\pm 0.3)\times 10^{-10}$ estimated by considering only stars with metallicity in the range where no scatter in excess of the observational errors is observed \cite{2010A&A...522A..26S}. Based on these analyses, we provide the first BBN constraints on CMB temperature evolution and expansion rate at very early times. We also study the primordial lithium problem~\cite{2011ARNPS..61...47F} using the non-parametric model of expansion rate, and find that the lithium abundance is consistent with relatively large deviations to the early expansion history around the onset of nuclear interactions.

We take yet another approach to reconstructing the early expansion history and allow for additional matter and radiation energy densities in the early universe, beyond the standard components of the universe. By comparing to BBN yield data, we find that only 15\% extra radiation energy density and 1\% extra matter energy density is allowed around the epoch of BBN. We then adopt a new and general analytic fitting formula for the temperature variation, capable of approximating the radiation temperature evolution in models where an entropy dump into radiation occurs in the early universe. This parameterization allows us to explore altered thermal histories independent of the specific microphysics model that would drive these deviations. We constrain the free parameters of this empirical model using BBN data, together with the measurements of the effective number of light species $N_\mathrm{eff}$ and $Y_\mathrm{p}$ from \textit{Planck} CMB anisotropy~\cite{2020A&A...641A...6P}. The resulting constraints can be mapped to the constraints on the new physics that may alter photon or standard neutrino temperature in nearly any particle model considered previously in the literature.

The organization of the paper is as follows. In Section \ref{sec:beyondSBBN}, we briefly review standard BBN physics, map out the redshift sensitivity of the light element abundances using Fisher martix analysis, and perform a non-parametric reconstruction of the radiation temperature evolution and expansion history in the early universe. In Section \ref{sec: param}, we introduce the empirical models to quantify deviations from the standard model of cosmology, and derive the constraints on the relevant parameters using BBN yields and CMB data. We summarize and discuss our findings in section \ref{sec:conclution}.

%%%%%%%%%%%%%%%%%%%%%%%%%%%%
\section{Non-parametric Approach}\label{sec:beyondSBBN}

%%%%%%%%%%%%%%%%%%%%%%%%%%%%
\subsection{Standard BBN}\label{sec:standard}

The standard BBN (SBBN) includes the $\Lambda$CDM model of cosmology and the SM of particle physics. In SBBN, the radiation energy density consists of photons and three flavors of light, left-handed neutrinos. The key free parameter that controls BBN in this case is the baryon density, parametrized by $\eta$, the ratio between the baryon number density $n_b$ and the photon number density $n_{\gamma}$ \cite{2012arXiv1208.0032S}. Using the measurements of the CMB temperature today, the present-epoch value of the baryon-to-photon ratio can be written as
\begin{equation}
    \eta_{10} = 10^{10}\frac{n_{b,0}}{n_{\gamma,0}} \approx 273.9 \Omega_{b,0}h^2 \ ,
\end{equation}
where $\Omega_{b,0}h^2$ is the density parameter for baryons, h is the Hubble constant, and the subscript ``0'' denotes present-day values. The factor of $10^{10}$ is a convenient scaling, since this ratio of the number densities is of the order of $10^{-10}$. The current observations place a stringent bound on $\eta_{10}$ (or, equivalently, $\eta$). For example, \textit{Planck} measurements imply that $\Omega_{b,0} h^2 = 0.02236\pm 0.00015$, at $68\%$ confidence level (CL) \cite{2020A&A...641A...6P}, corresponding to $\eta_{10} = 6.12\pm 0.04$. Within such a small uncertainty, the variation of baryon density does not significantly affect the primordial abundance of light elements. Therefore, we fix the value of $\eta_{10}$ to 6.12 in our analysis.

%%%%%%%%%%%%
\begin{figure}
\includegraphics[width=0.44\textwidth]{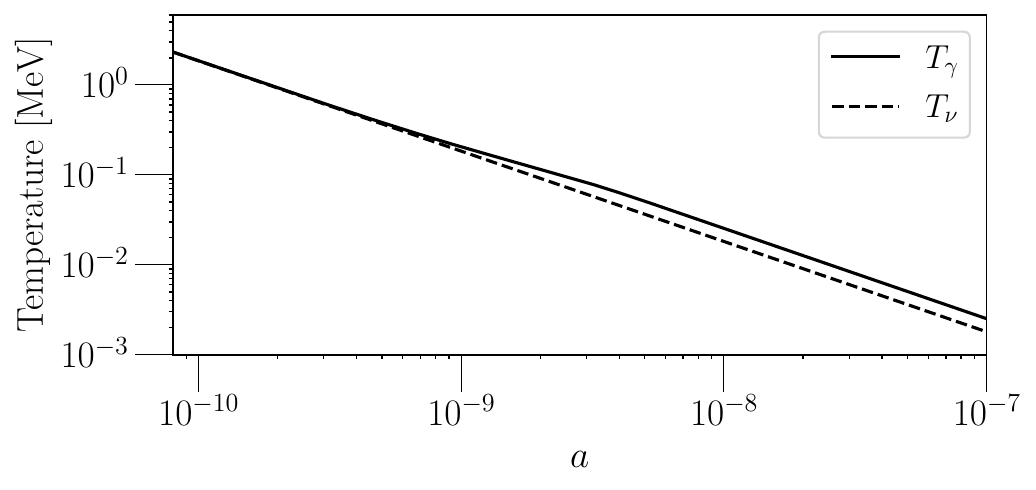}
\caption{The evolution of photon temperature $T_{\gamma}$ (solid curve) and neutrino temperature $T_{\nu}$ (dashed curve) as a function of the scale factor $a$, in the standard BBN scenario.}\label{fig:Temp}
\end{figure}
%%%%%%%%%%%%
%%%%%%%%%%%%
\begin{figure}
\includegraphics[width=0.44\textwidth]{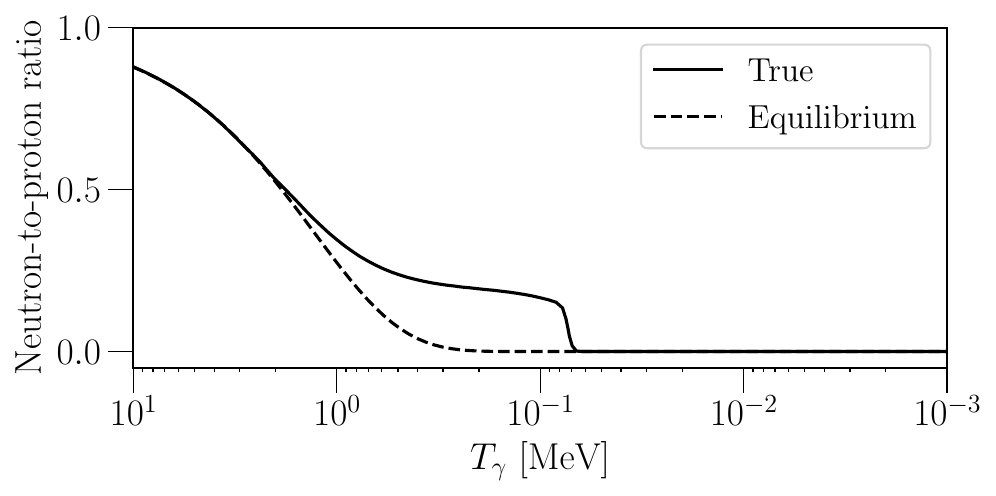}
\caption{The evolution of the neutron-to-proton number-density ratio $n/p$ as  a function of photon temperature $T_{\gamma}$. The solid curve indicates the ``true'' ratio in the standard BBN case, and is generated by \texttt{AlterBBN} code; the dashed curve indicates the equilibrium ratio following $e^{-\Delta m/T_{\gamma}}$, where $\Delta m$ is the neutron-proton mass difference.}\label{fig:npratio}
\end{figure}
%%%%%%%%%%%%

In order to build intuition for the effects of the deviations to standard cosmology on BBN, we first review some of the key points of the BBN process. At the beginning of this process, the universe was radiation-dominated, containing electrons and positrons $e^{\pm}$, photons $\gamma$, three neutrino species $\nu$, and a small number of protons $p$ and neutrons $n$; all the species are assumed to be in thermal equilibrium with each other. When the temperature drops below $2-3$ MeV (corresponding to initial time $a_\mathrm{ini}$, as shown in Figure \ref{fig:Temp}), the interaction rates of neutrinos become slower than the expansion rate, and the neutrinos effectively decouple from the photons and $e^{\pm}$ \cite{1992NuPhB.374..392E, 2002PhR...370..333D, 2002PhRvD..65h3006H}. However, the electron neutrinos $\nu_{e}$ continue to interact with the nucleons via the charged-current weak interactions, until the temperature drops below $\sim 0.8$ MeV (at $3 \times 10^{-10}$). The two-body interactions among $n$, $p$, $e^{\pm}$, and $\nu_{e} (\bar{\nu}_{e})$ continue to influence the ratio of neutrons to protons $n/p$, although not rapidly enough to allow $n/p$ to track its equilibrium value \cite{1996RPPh...59.1493S,2007ARNPS..57..463S} (see Figure \ref{fig:npratio}). As a result, $n/p$ continues to decrease from $\sim 1/6$ at freeze out to $\sim 1/7$, at the onset of the next phase of BBN: the formation of nuclei at $\lesssim 0.08$ MeV (corresponding to $a \gtrsim 4 \times 10^{-9}$). This phase involves various nuclear interactions between $n, p$, $D$ and other light nuclei, and ultimately results in the production of $D$, ${}^3\!He$, ${}^4\!He$ and ${}^7\!Li$. These reactions occur until their rates drop below the expansion rate.

%%%%%%%%%%%%%%%%%%%%%%%%%%%%
\subsection{Modifications to the Expansion History}\label{sec:expansion}

At the time of BBN, the universe is dominated by radiation, with the total energy density $\rho_{R}$, such that
\begin{equation}
    H^2(a) = \frac{8\pi G}{3}\rho_R(a)\ ,
\end{equation}
where $H$ is the Hubble parameter, $a$ is the scale factor, and $G$ is the Newton constant. Presence of new particles may lead to extra energy density, such that
\begin{equation}
    H(a) = H_\mathrm{fid}(a)[1+\delta_H (a)]\ , \label{eq:H} 
\end{equation}
where we use variable $\delta_H$ to capture any variation from the fiducial (standard) expansion history, following Ref.~\cite{2012PhRvD..86l3504S}. This leads to modifications to the predictions of SBBN. For example, a faster expansion causes the rate of the weak interaction that interconverts neutrons and protons to fall out of equilibrium at a higher temperature. Since there are more neutrons relative to protons in equilibrium at higher temperatures, this leads to an increase in the BBN yield of helium-4. Meanwhile, the increased expansion rate leaves less time for deuterium destruction, and thus also raises its relic abundance.

To analyze the constraining power of the BBN yields on $\delta_H$, we begin by writing it as a linear combination in an orthogonal bin basis,
\begin{equation}
    \delta_H (a) = \sum_i  \delta_{H,i} b_i\ , 
    \label{eq:deltaH} 
\end{equation}
where  
\begin{equation}
    b_i  =  \left[\frac{1}{1+e^{(\ln a-\ln a_{i+1})/\tau}}-\frac{1}{1+e^{(\ln a-\ln a_i)/\tau}}\right]\ . 
    \label{eq:bi}
\end{equation}
Within bin $i$, $\delta_H = \delta_{H,i}$ and far from any other bins $\delta_H=0$. The bin edges are slightly smoothed, using a Gaussian smoothing of width $\tau$ times the bin width, to prevent infinite derivatives. We can then consider perturbations to $H$ in bins of $a$, which is a model independent description. 

To quantify the impact of the deviations $\delta_H$ on BBN, we implement Eqs.~(\ref{eq:H}-\ref{eq:bi}) into the publicly-available \texttt{AlterBBN} code \cite{Jenssen2016NewAA, 2018arXiv180611095A}, which enables high-accuracy BBN predictions. We use 30 bins over the range of $a = [8\times 10^{-11}, 10^{-7}]$, logarithmically spaced, where $8\times 10^{-11}$ is the initial scale factor $a_\mathrm{ini}$, set in \texttt{AlterBBN} code, and $a = 10^{-7}$ is far after the BBN process ends. 

To estimate the relative significance of the deviations $\delta_H$ at different redshifts, we carry out a Fisher matrix calculation. The Fisher matrix elements are given by
\begin{equation}
    F_{ij} = \left(\frac{\partial{\mathbf{A}}}{\partial{p_{i}}}\right)^\mathrm{T}  \textbf{Cov}^{-1} \frac{\partial{\mathbf{A}}}{\partial{p_{j}}}
    \ ,
    \label{eq:def_FIM}
\end{equation}
where $\mathbf{A}$ is a vector of the BBN observables, namely the primordial abundances of light elements. The covariance matrix $\textbf{Cov}$ is given by the measured uncertainties from BBN observations reported in Ref.~\cite{2020PTEP.2020h3C01P}, with $\sigma(Y_\mathrm{p})=0.003$, $\sigma(Y_\mathrm{D})=0.025$, $\sigma({}^3\!He/H)=0.2\times 10^{-5}$ and $\sigma({}^7\!Li/H)=0.3\times 10^{-10}$. In our analysis, besides the observational uncertainties, we also account for the theoretical uncertainty which is related to the uncertainties on neutron
lifetime and various nuclear reaction rates (see details in Appendix~\ref{appendix.A}). The parameter set \{$p_i$\} contains the $N$-bin-expansion parameters $\delta_{H,i}$. The uncertainties on each $\delta_{H,i}$ are given by the square root of the respective diagonal element of the inverse of the Fisher matrix, $\sigma({\delta_{H,i}}) = \sqrt{(F^{-1})_{ii}}$.

%%%%%%%%%%%%
\begin{figure}
\includegraphics[width=0.46\textwidth]{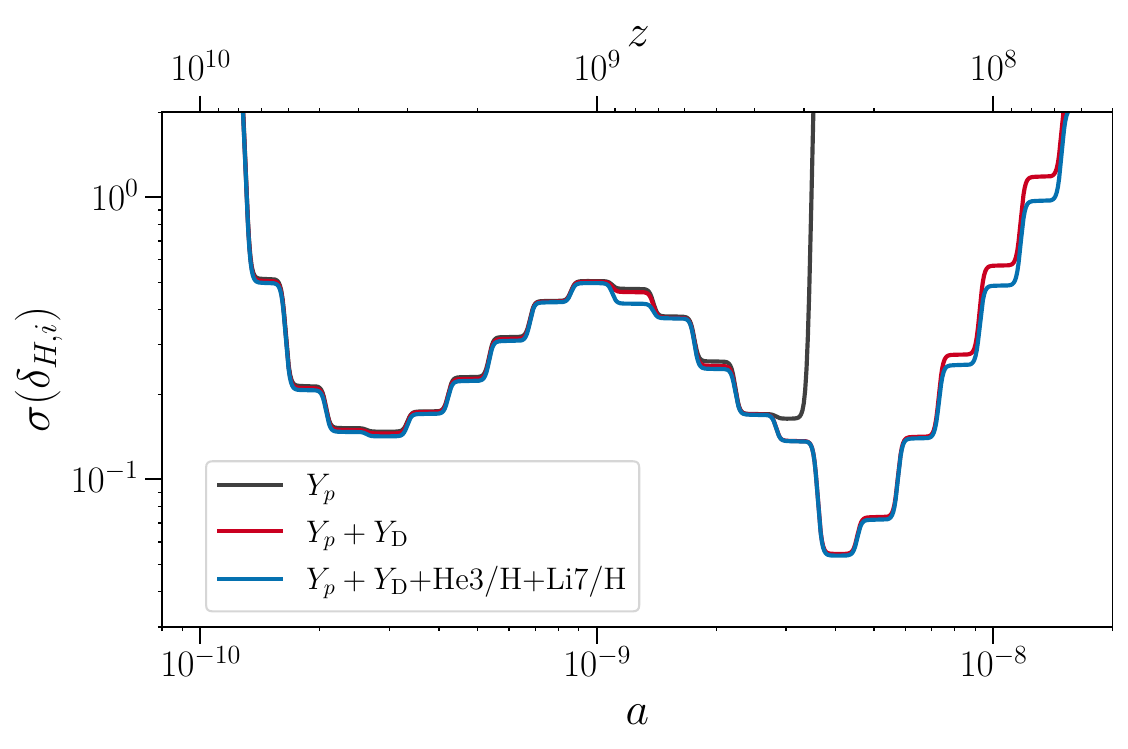}
\caption{The projected uncertainties on the expansion rate derived from light element abundance measurements. We show deviations $\sigma(\delta_{H,i})$ from the standard rate, per bin in scale factor, derived from Fisher matrix analysis, using three combinations of the BBN measurements: helium-4 abundance alone $Y_\mathrm{p}$; helium-4 and deuterium abundances $ Y_\mathrm{p}+Y_\mathrm{D}$; and all the available observed abundances to date $Y_\mathrm{p}+Y_\mathrm{D}+{}^3\!He/H+{}^7\!Li/H$. We use 30 bins in the range $a = [8\times 10^{-11}, 10^{-7}]$, logarithmically spaced. }\label{fig:fisher_H}
\end{figure}
%%%%%%%%%%%%

%%%%%%%%%%%%
\begin{figure*}
\includegraphics[width=0.46\textwidth]{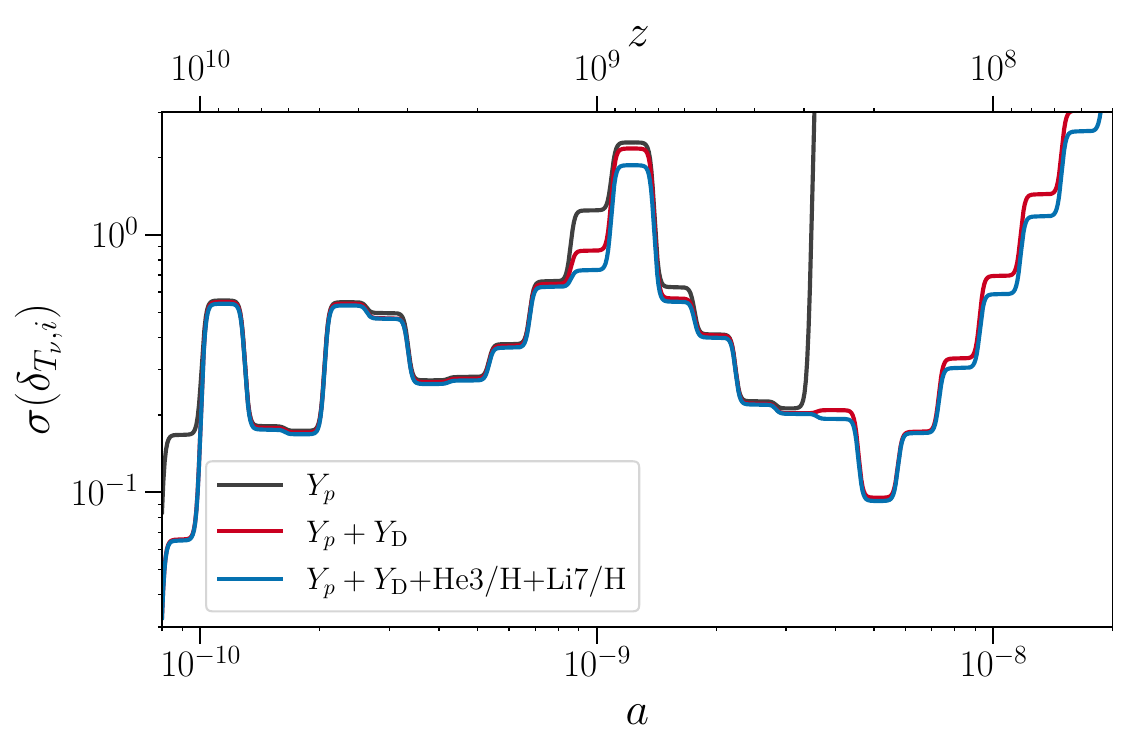}\ \ \ \ \ \ 
\includegraphics[width=0.46\textwidth]{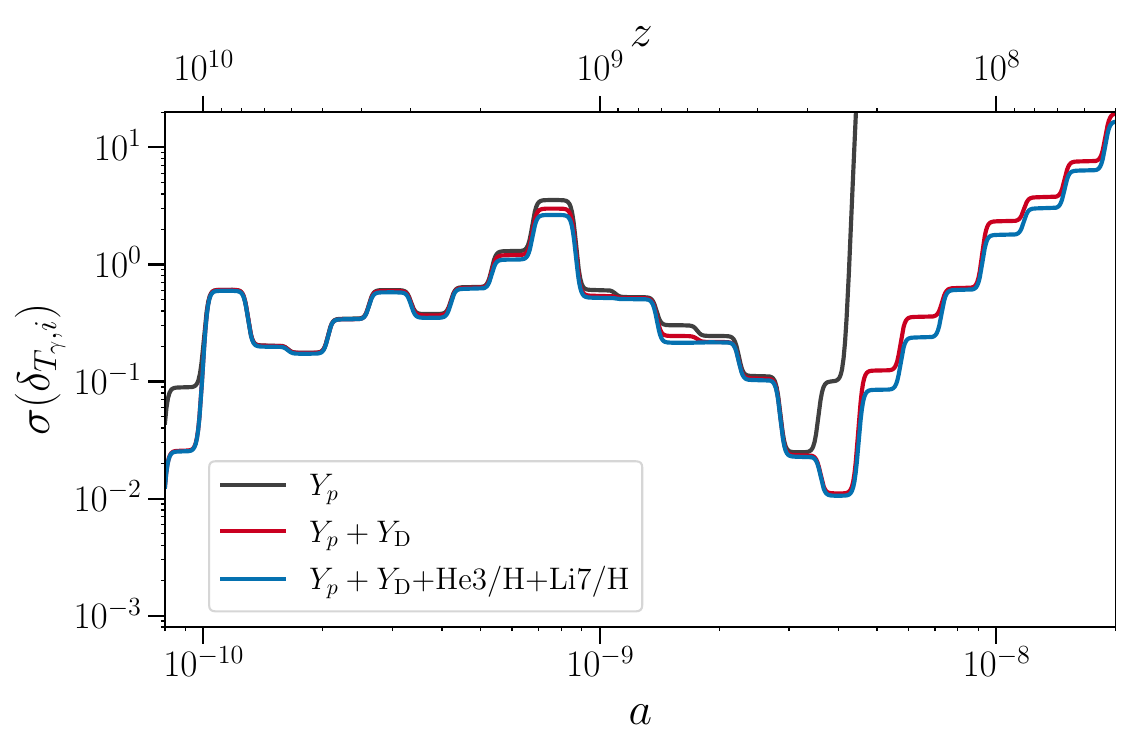}
\caption{Similar to Figure \ref{fig:fisher_H}, but showing the uncertainties on neutrino temperature deviations $\sigma(\delta_{T_{\nu},i})$ (left panel), and photon temperature deviations $\sigma(\delta_{T_{\gamma},i})$ (right panel).}\label{fig:fisher_Tem}
\end{figure*}
%%%%%%%%%%%%

Figure \ref{fig:fisher_H} shows the result of our analysis, in the form of the uncertainties on the expansion bin parameters, given a set of BBN yield measurements. Here we consider three combinations of the yield data: ${}^4\!He$ abundance alone $\mathbf{A} = (Y_\mathrm{p})$; ${}^4\!He$ and $D$ abundances $\mathbf{A} =  (Y_\mathrm{p}, Y_\mathrm{D} )$; and all the available observed abundances to date $\mathbf{A} = ( Y_\mathrm{p}, Y_\mathrm{D}, {}^3\!He/H, {}^7\!Li/H )$, including the abundances of ${}^3\!He$ and ${}^7\!Li$. One can see that the abundance of ${}^4\!He$ is only sensitive to the expansion rate at $a\lesssim 4 \times 10^{-9}$, which is because $Y_\mathrm{p}$ is mainly determined by the neutron-to-proton ratio at the onset of nuclear interactions. It can be seen that deuterium abundances $Y_\mathrm{D}$ is sensitive to the expansion history after $a\sim 4 \times 10^{-9}$, and its sensitivity diminishes with time. Including the measurements of ${}^3\!He$ and ${}^7\!Li$ has no significant effects on the uncertainties. We notice the maximum of the information content are around $n/p$ decoupling ($a\sim 3 \times 10^{-10}$) and the onset of nuclear interactions ($a\sim 4 \times 10^{-9}$). With $Y_\mathrm{p}$ alone one can achieve 10\%-level constraints on the expansion rate around those two epochs; with all four observed abundances we obtain percent-level constraints near the latter epoch. 

\subsection{Modifications to the Thermal History}\label{sec:temperature}

In addition to the expansion rate, we next consider the entropy content of the early universe, which can also feature modifications, e.g. from the light particles that annihilate during BBN and heat photons relative to the neutrinos, or vice versa. In the case of light particles annihilating into neutrinos, the temperature of neutrinos $T_{\nu}$ can increase relative to photons, if the annihilation occurs after neutrino decoupling. In addition to contributing directly to the relativistic energy density, this process would also speed up the weak rates interconverting neutrons and protons. Theses two effects add up and increase the abundances of ${}^4\!He$ and $D$, compared to that in the SBBN.
%The opposite happens if the annihilation increases the temperature of photons. 
On the other hand, if the annihilation increases the temperature of photons, the effect on BBN yields depends on the exact timing of the annihilation~\cite{2022JCAP...07..002A}. We consider the two scenarios separately.

We now model the temperature of photons and neutrinos non-parametrically, using a similar approach to that employed for the expansion history. We consider deviations from the standard thermal history over 30 bins in the range $a = [8\times 10^{-11}, 10^{-7}]$, logarithmically spaced.
Analogous to $\delta_{H}$, we parameterize the variations from the fiducial temperature as
\begin{equation}
    \delta_{T_{x}} (a) = \sum_i  \delta_{T_{x},i} b_i \ ,
\end{equation}
where $x\in \{\nu, \gamma\}$, for neutrinos and photons, respectively.

We again carry out a Fisher matrix calculation and present our results in Figures \ref{fig:fisher_Tem}.
With all the observed abundances, we find a percent-level uncertainty at $a_\mathrm{ini}$. As expected, the temperature around the $n/p$ decoupling and the onset of nuclear interactions are also very important to the BBN process. We find that BBN observations probe the scale factor range $a \lesssim 10^{-8}$, and the maximum of the information content is at the time of neutrino decoupling, and around the onset of nuclear interactions, where we project percent-level constraints on the radiation temperature.

%%%%%%%%%%%%%%%%%%%%%%%%%%%%
\subsection{Non-parametric Reconstruction}\label{subsec:non-param}

%%%%%%%%%%%%
\begin{figure*}
\includegraphics[width=0.9\textwidth]{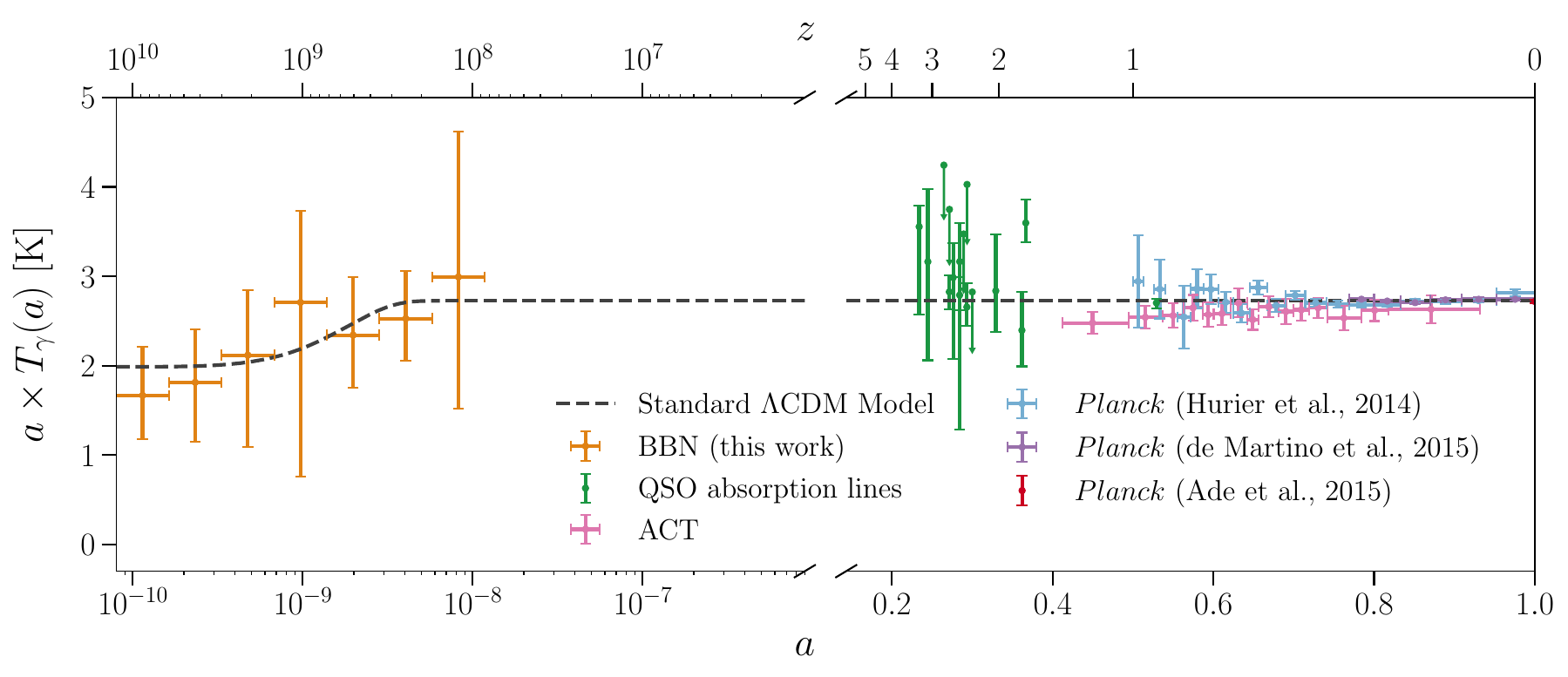}
\caption{Reconstruction of the radiation temperature from BBN yield measurements, including  $Y_\mathrm{p}$ and $Y_\mathrm{D}$. The error bar of each point is the 68\% CL interval in each redshift bin. We also show previous results, derived from quasar absorption line studies (green) \cite{2013A&A...551A.109M,2020AstL...46..715K}, from the analyses of Sunyaev-Zel'dovich galaxy clusters from the Atacama Cosmology Telescope (pink) \cite{2021ApJ...922..136L}, from two independent analyses of \textit{Planck} clusters (blue and purple) \cite{2014A&A...561A.143H,2015ApJ...808..128D}, and from \textit{Planck} CMB data (red) \cite{2016A&A...594A..13P}. The dashed line marks the standard evolution of $T_\gamma(a)$.}\label{fig:CMB_Tem}
\end{figure*}
%%%%%%%%%%%%

Given the cosmological importance of the photon temperature, we now turn to reconstructing this quantity in the early universe, using a non-parametric approach presented in previous sections. Guided by the Fisher matrix analysis in Section \ref{sec:temperature}, we use 10 redshift bins that have the highest impact on BBN yields\footnote{We have investigated using different numbers of redshift bins and found that this number captures the information near the key points of the BBN process without losing the sensitivity to the observables: bin 1 is near the neutrino decoupling; bins 2 and 3 correspond to the epochs right before and after $n/p$ decoupling; bins 5 and 6 correspond to the epochs right before and after the onset of BBN; bin 7 corresponds to the BBN freeze-out; and bins 8--10 capture the physics after the end of BBN.}.

In this model, the free parameters $\{p_i\}$ are the 10 redshift-bin values of the temperature variations $\delta_{T_{\gamma},i}$. From previous sections, we know that the constraining power from BBN yields mainly comes from $Y_\mathrm{p}+Y_\mathrm{D}$. Thus, we only consider the measurements of these two elements in our main analysis. We discuss the effects of including measurements of ${}^3\!He$ and ${}^7\!Li$ on our results, in the following section. The observed abundances of ${}^4\!He$ and $D$ are $Y_\mathrm{p}=0.245\pm{0.003}$, $Y_\mathrm{D}=2.547\pm{0.025}$~\cite{2020PTEP.2020h3C01P}. To determine the values of $\{ p_i\}$ that are consistent with the primordial abundance data, we construct a chis-quared statistic $\chi^2_\text{BBN}$ that depends on the model parameters (see details in Appendix~\ref{appendix.A}), and perform a MCMC analysis using the publicly-available \texttt{emcee} code~\cite{2013PASP..125..306F}. We utilize its \texttt{mcmc} sampler, and employ the convergence criterion $R-1 = 0.01$, where $R$ is the Gelman-Rubin threshold \cite{1992StaSc...7..457G}. 

Figure \ref{fig:CMB_Tem} captures our key result, showing the mean values and 68\% CL errors for the temperature variations. Here we only show the results in the first 7 bins, since the last 3 bins are at $a\gtrsim 10^{-8}$, where the deviations can not be tightly constrained by the data (see the full probability distribution for the 10 bin parameters in Appendix \ref{appendix.B}). We also show the previous measurements from quasar absorption line studies \cite{2013A&A...551A.109M,2020AstL...46..715K}, from the analyses of Sunyaev-Zel'dovich galaxy clusters from the Atacama Cosmology Telescope \cite{2021ApJ...922..136L}, from two independent analyses of \textit{Planck} clusters \cite{2014A&A...561A.143H,2015ApJ...808..128D}, and from \textit{Planck} CMB data \cite{2016A&A...594A..13P}. This figure represents the best current model-independent reconstruction of the CMB temperature at high and low redshifts. The intermediate redshift range may be filled by other cosmological measurements, which we have not considered in this study.

\begin{figure*}
\includegraphics[width=0.8\textwidth]{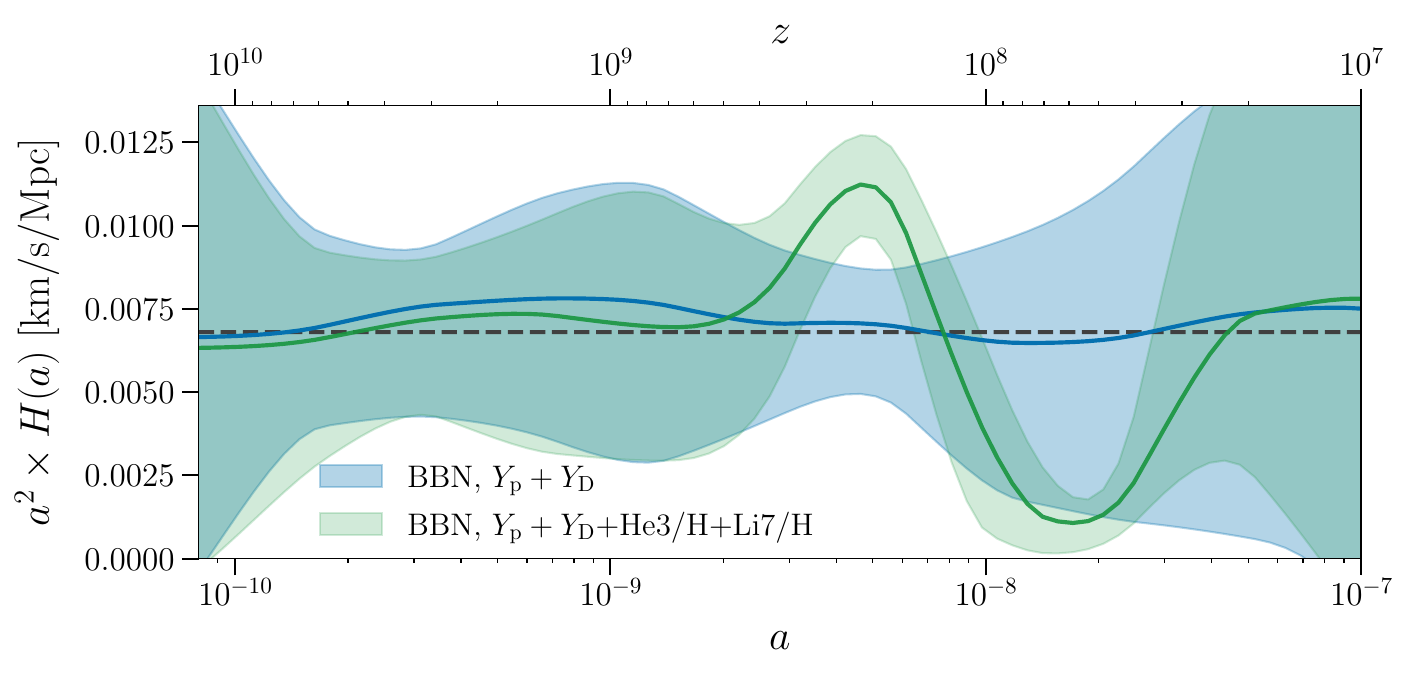}
\caption{Reconstruction of the expansion rate from BBN yield measurements. The blue region indicates the deviations that allowed at 68\% CL by the measurements of $Y_\mathrm{p}$ and $Y_\mathrm{D}$, and the blue solid line shows the mean value of the evolution; The green line and region correspond to the reconstruction from BBN data that includes all the available observed abundances $( Y_\mathrm{p}, Y_\mathrm{D}, {}^3\!He/H, {}^7\!Li/H )$; The dashed line marks the standard evolution of $H(a)$.}\label{fig:expansion}
\end{figure*}

Next, we repeat an analogous procedure to measuring the early-universe expansion history. The reconstructed model of the expansion rate with mean value (blue solid line) and 68\% CL range (blue region) is shown in Figure \ref{fig:expansion}, where the curve and edges of the region are smoothed. We can see the standard $\Lambda$CDM fits the observations well. The tightest constraints on $H(a)$ variations are around $n/p$ decoupling and the onset of nuclear interactions, which are consistent with the previous analysis in Section \ref{sec:expansion}.

%%%%%%%%%%%%
%%%%%%%%%%%%

%%%%%%%%%%%%%%%%%%%%%%%%%%%%
\subsection{The Primordial Lithium Problem}\label{subsec:Li_problem}

While the primordial abundances adopted from SBBN are in very good agreement with the observed values for $D$, ${}^4\!He$ and ${}^3\!He$\footnote{In the case of ${}^3\!He$, the only data available come from the Solar System and from the solar-metallicity HII regions in the Galaxy~\cite{2002Natur.415...54B,2018AJ....156..280B}. Therefore, inferring the primordial ${}^3\!He$ abundance may be considered less reliable for the BBN study purposes.}, the SBBN-predicted abundance of ${}^7\!Li$ is factor of 3--4 higher than its observed value; this discrepancy is known as the primordial lithium problem~\cite{2011ARNPS..61...47F}.
%For instance, it is found ${}^7\!Li/H = (5.46 \pm 0.43)\times 10^{-10}$ in SBBN using \texttt{AlterBBN} code, which a factor of 3.4 higher than the observed value ${}^7\!Li/H = (1.6 \pm 0.3)\times 10^{-10}$~\cite{2020PTEP.2020h3C01P} and has a discrepancy up to $\sim 10\sigma$ CL. 
The lithium problem remains an unresolved issue and a variety of suggestions have been proposed to remedy this discrepancy~\cite{2009NJPh...11j5028J,2011ARNPS..61...47F,2004PhRvD..69l3519C}. Some of the most popular explanations include new physics, during or soon after BBN. 

For this reason, in our non-parametric approach, we first reconstruct the early expansion history using the abundances of $D$ and ${}^4\!He$ only, as our baseline result; we then show how this result changes when we include ${}^3\!He$ and ${}^7\!Li$. In Figure~\ref{fig:expansion}, the shaded regions indicate deviations consistent with a given set of primordial abundance measurements at 68\% CL, and the solid lines corresponds to the mean value in a given analysis. Comparing the results from the measurements of $Y_\mathrm{p}$ and $Y_\mathrm{D}$ only to the results that include primordial lithium measurements, we note that the reconstruction changes significantly around and soon after the onset of nuclear interactions, showing large deviations to the SBBN evolution (dashed black line).

The expansion rate that exceeds the SBBN evolution around the onset of nuclear interactions is required by the lithium data, which can be understood as follows. Most of the primordial ${}^7\!Li$ comes from the decay of beryllium-7 ${}^7\!Be$, and since ${}^7\!Be$ is still being produced at the end of BBN, a faster expansion leads to a decrease in its abundance, and consequently, a lower relic ${}^7\!Li$ abundance---bringing measurements into better consistency with the BBN predictions. We note that the slower expansion seen afterwards has an opposite but relatively subdominant effect on ${}^7\!Li$ abundance; however, it reduces the spike in the abundances of $D$ and ${}^3\!He$ caused by the faster expansion, and brings them back into agreement with the measurements. The large deviations to the early expansion history around and soon after the onset of nuclear interactions are therefore a viable solution to the lithium problem, although their physical driving mechanism remains unclear.

%%%%%%%%%%%%%%%%%%%%
\section{Empirical Model}\label{sec: param}

In this section, we concentrate on another approach to parameterize and reconstruct the evolution of the expansion rate and the thermal history, avoiding assumptions about the specific particle content of the universe. Namely, we consider physically-motivated empirical models for the energy density and radiation temperature, respectively. In both cases, our empirical parameterization is flexible enough to capture a wide range of beyond-standard-BBN physics considered in the literature \cite{2012CoPhC.183.1822A, 2014PhRvD..89h3508N,2015PhRvD..91h3505N,2019JCAP...02..007E,2020JCAP...01..004S,2022PhRvL.129b1302G,2022JCAP...07..002A}, as discussed below.

%%%%%%%%%%%%%%%%%%%%%%%%%%%%
\subsection{Expansion Rate during BBN}

We first allow for additional matter and radiation energy density $\rho_\mathrm{D}(T)$, beyond the standard cosmological model, such that~\cite{2008PhLB..669...46A}
\begin{equation}
    \rho_\mathrm{D}(T)=\rho_\mathrm{R}(T_0)[\kappa_m (\frac{T}{T_0})^3+\kappa_{\gamma} (\frac{T}{T_0})^4],
    \label{eq.paramH}
\end{equation}
where $T$ represents radiation temperature; $\kappa_m$ is the ratio of matter-like energy density over the total energy density at the BBN temperature $T_0=1$ MeV,  $\kappa_{\gamma}$ is the ratio of relativistic; both $\kappa_m$ and $\kappa_{\gamma}$ are free parameters of this model.
In this case, the additional energy density leads to a change in the expansion rate, as captured by the Friedmann equation,
\begin{equation}
    H^2=\frac{8\pi G}{3}(\rho_\mathrm{R}+\rho_\mathrm{D}),
\end{equation}
where $\rho_\mathrm{R} \sim \rho_\mathrm{tot}$ represents the total radiation energy density in the SBBN scenario. We implement this model into the \texttt{AlterBBN} code to quantify its impact on the BBN yields. 

To place the BBN bounds on this model, we carry out an MCMC analysis, as in Section~\ref{subsec:non-param} and sample the posterior distribution of $\{\kappa_{\gamma}, \kappa_m\}$. For each parameter, we employ broad priors $\kappa_{\gamma} \in [0, 0.99]$ and $\kappa_{m} \in [0, 0.99]$. We only consider the measurements of $Y_\mathrm{p}$ and $Y_\mathrm{D}$ here, for the same reasons in Section~\ref{subsec:non-param}.

%%%%%%%%%%%%
\begin{figure}
\includegraphics[width=0.40\textwidth]{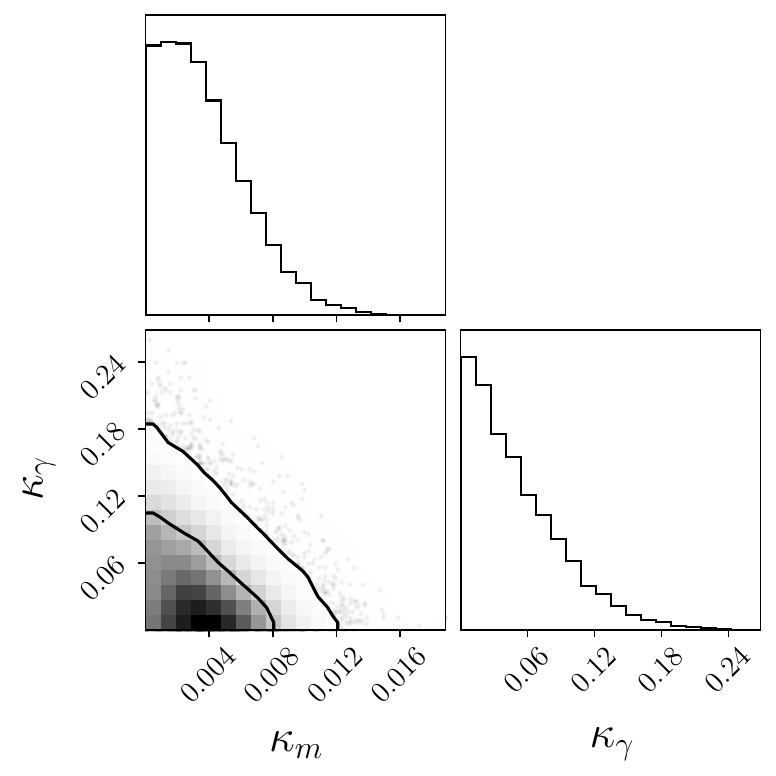}
\caption{The posterior probability distribution for the expansion rate parameters. We show the 68\% and 95\% CL contours, obtained from measurements of $Y_\mathrm{p}$ and $Y_\mathrm{D}$.  The one-dimensional, marginalized posteriors are shown at the top of each column.}\label{fig:Constraints_H}
\end{figure}
%%%%%%%%%%%%

\begin{figure}
\includegraphics[width=0.46\textwidth]{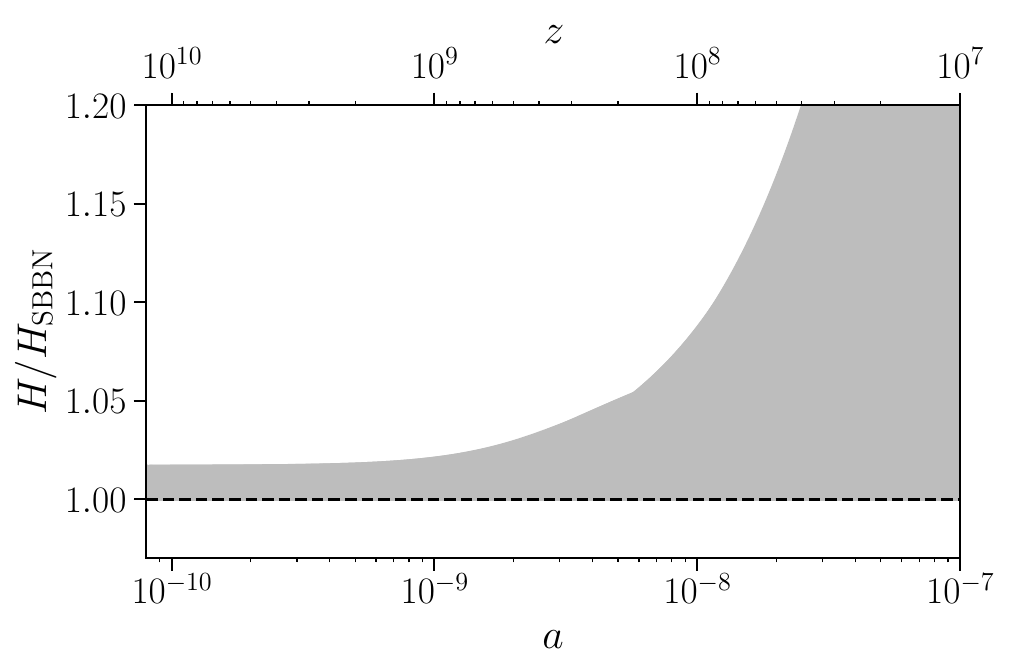}
\caption{The gray area shows the $95\%$ CL region for the expansion rate, as inferred from the measurements of BBN. }\label{fig:hubble-empirical}
\end{figure}

The resulting posterior probability distributions are shown in Figure \ref{fig:Constraints_H}. As expected, there is a prominent (negative) degeneracy between $\kappa_{\gamma}$ and $\kappa_{m}$. We find that the current measurements of the primordial helium and deuterium abundances are able to place a upper bound on the additional matter energy density fraction of $\kappa_{m}<0.01$ at 95\% CL, given by its marginalized posterior distribution, which is much tighter as compared with the 95\% CL bounds on the additional radiation component contribution $\kappa_{\gamma}<0.15$. We illustrate the derived uncertainties on these parameters in the form of allowed error band on the Hubble parameter in Figure \ref{fig:hubble-empirical}. The gray band indicates the region allowed by BBN measurements at $95\%$ CL.

%%%%%%%%%%%%
%%%%%%%%%%%%

%%%%%%%%%%%%%%%%%%%%%%%%%%%%
\subsection{Temperature during BBN}\label{subsec:temp}

If the additional energy-density components interact with the rest of the plasma during BBN, they may deposit entropy into other species and affect radiation temperature in the early universe, leaving imprints on the BBN abundance of light elements. We find that the contributions to the radiation temperature from an entropy dump in various models considered in previous literature, involving annihilating DM, can be approximated by the following fitting function,
\begin{equation}
    \frac{T_x}{T_\mathrm{SBBN}}(a) = 1+ \delta_x \{1-[1+(\alpha_x \times (a-a_\mathrm{ini}))^{\beta_x}]^{\gamma_x} \},
    \label{eq.paramT}
\end{equation}
where $\alpha_x$, $\beta_x$, $\gamma_x$ and $\delta_x$ are free parameters, $a_\mathrm{ini} = 8\times 10^{-11}$ is the initial scale factor, and $x\in\{\nu, \gamma\}$ stands for the neutrino and photon fluids, respectively. The value of $\alpha$ specifies the general time range for onset of the entropy dump; $\beta$ and $\gamma$ control the speed of the annihilation process; and $\delta$ determines the total energy transferred in the process.

%%%%%%%%%%%%
\begin{figure*}
\includegraphics[width=0.46\textwidth]{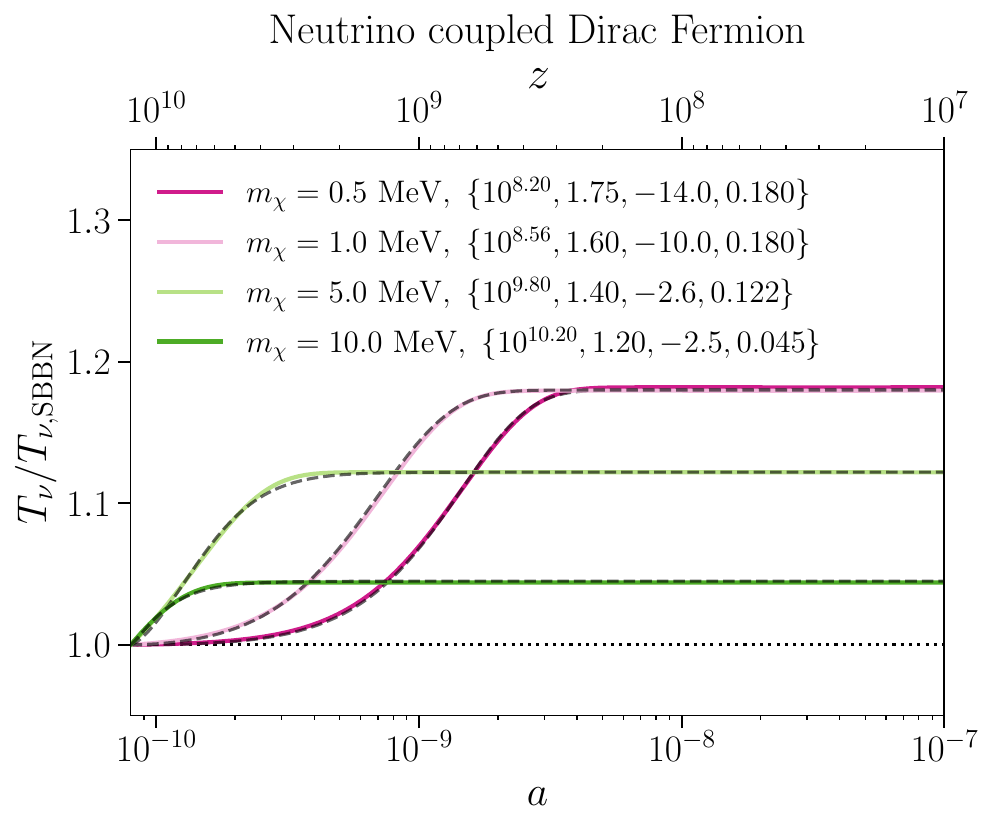}
\includegraphics[width=0.46\textwidth]{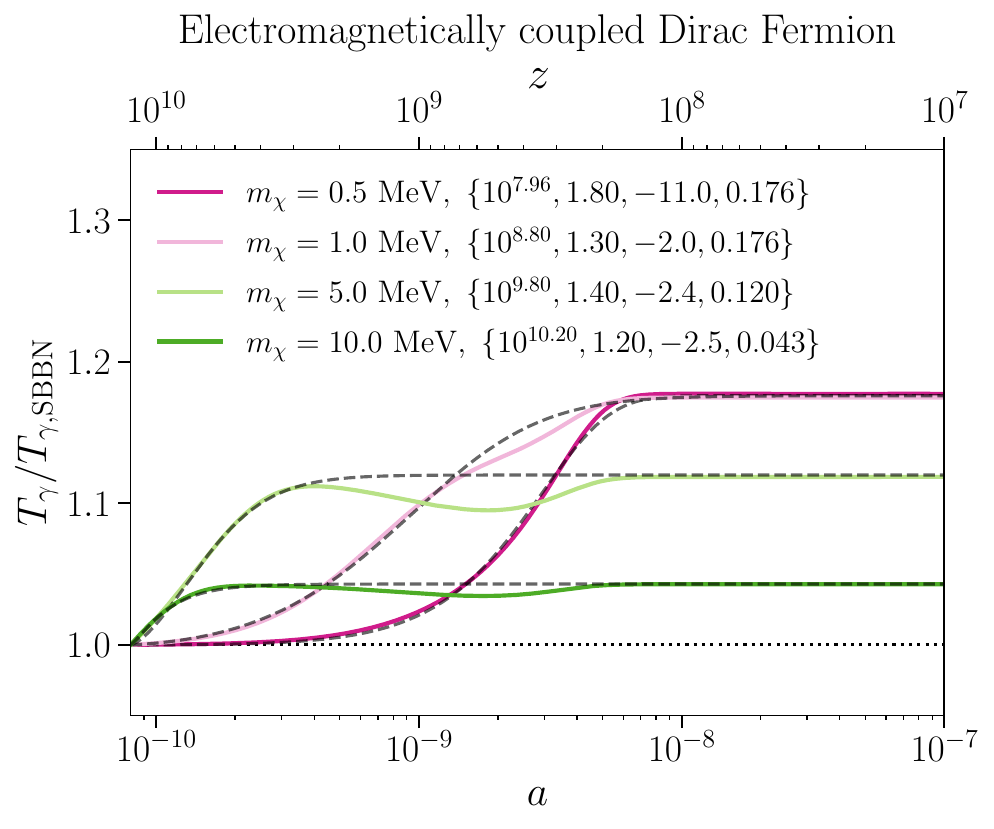}
\caption{The ratio of the radiation temperature in a cosmology featuring light thermally-coupled Dirac Fermion DM model, with a particle mass $m_{\chi}$ shown in the legend, to the standard BBN scenario. The case of DM coupling electromagnetically is shown in the right panel, and DM coupled to the SM neutrinos is in the left panel. The colored solid curves are obtained using \texttt{AlterBBN} code. The dashed gray curves are obtained from the empirical model shown in Eq.~(\ref{eq.paramT}), and the corresponding parameters are listed in the bracket: $\{\alpha_x, \beta_x, \gamma_x, \delta_x\}$. We find that the empirical model has sufficient flexibility to accurately capture the rande of $T/T_\mathrm{SBBN}$ behaviors occurring in this set of DM scenarios.}\label{fig:fitting}
\end{figure*}
%%%%%%%%%%%%
We next check that our model has sufficient flexibility to capture known specific particle models that modify thermal history during BBN. As a broad set of examples, we focus on light thermal-relic DM with a mass smaller than $20$ MeV, which annihilates soon after neutrino decoupling, and becomes non-relativistic around the time of BBN \cite{2014PhRvD..89h3508N, 2015PhRvD..91h3505N, 2022JCAP...07..002A}. In this model, DM will dump entropy into existing radiation components, altering their temperatures and affecting the BBN yields as a result. With appropriate choices of $\alpha_x$, $\beta_x$, $\gamma_x$ and $\delta_x$ values, the fitting formula in Eq.~(\ref{eq.paramT}) can reproduce the temperature evolution in each case of DM annihilation scenarios, as show in Figure \ref{fig:fitting}. Specifically, the annihilation process leads to a deviation from the standard temperature $T_\mathrm{SBBN}$, depending on the type of DM particle and the type of coupling to the SM \cite{2022JCAP...07..002A}. In Figure \ref{fig:fitting}, we show the ratio of the temperature for a cosmology featuring light thermally coupled Dirac Fermion DM model to that of the standard BBN, for the case where DM annihilates into the SM neutrinos (left pannel) and for electromagnetically-coupled DM (right pannel). DM annihilating into photons can heat photons, while DM annihilating to neutrinos heats up neutrinos relative to photons. The resulting ${T}/{T_\mathrm{SBBN}}$ at different DM masses (colored solid curves) is obtained from \texttt{AlterBBN} code, while the corresponding fits of our empirical model $\{\alpha_x$, $\beta_x$, $\gamma_x$, $\delta_x\}$ (dashed curves) are obtained from Eq.~(\ref{eq.paramT}), and present a good fit to the range of annihilating DM scenarios considered here.

%%%%%%%%%%%%
\begin{figure}
\includegraphics[width=0.38\textwidth]{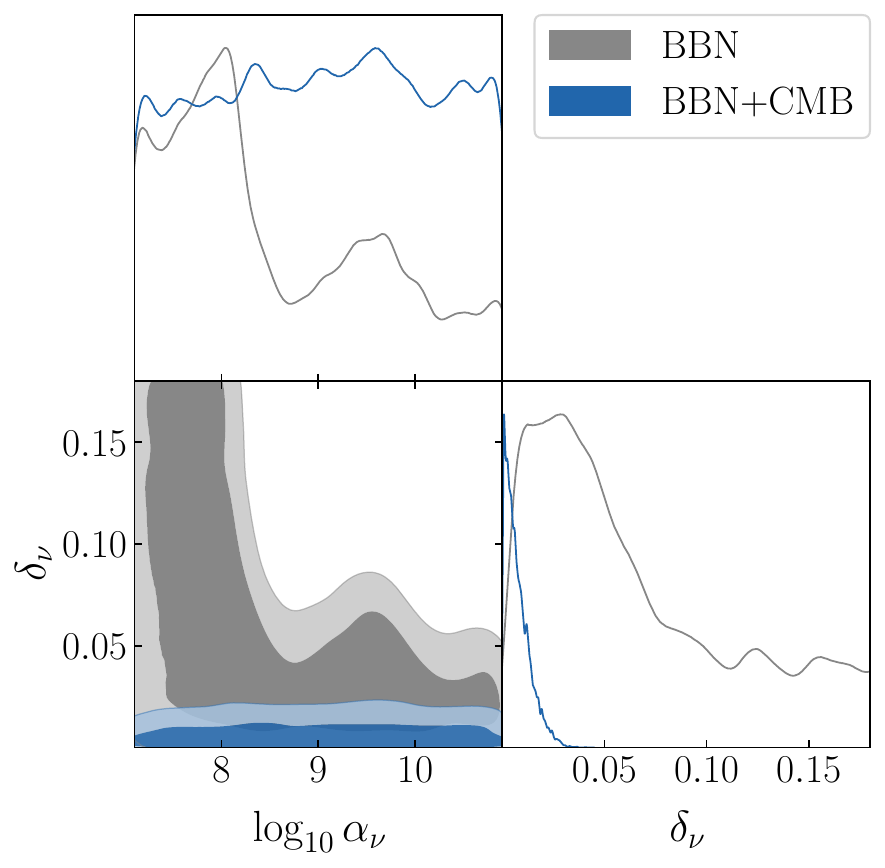}
\caption{The posterior probability distribution of parameters $\{\log_{10}\alpha_\nu, \delta_\nu \}$ for the scenario in which neutrino temperature deviates from its standard-model predictions. We show the 68\% and 95\% CL contours, obtained from the BBN measurements of $Y_\mathrm{p}$ and $Y_\mathrm{D}$, together with the CMB measurements of $N_\mathrm{eff}$ and $Y_\mathrm{p}$. The one-dimensional, marginalized posteriors are shown at the top of each column.}\label{fig:Constraints_Tnu}
\end{figure}
%%%%%%%%%%%%
%%%%%%%%%%%%
\begin{figure}
\includegraphics[width=0.38\textwidth]{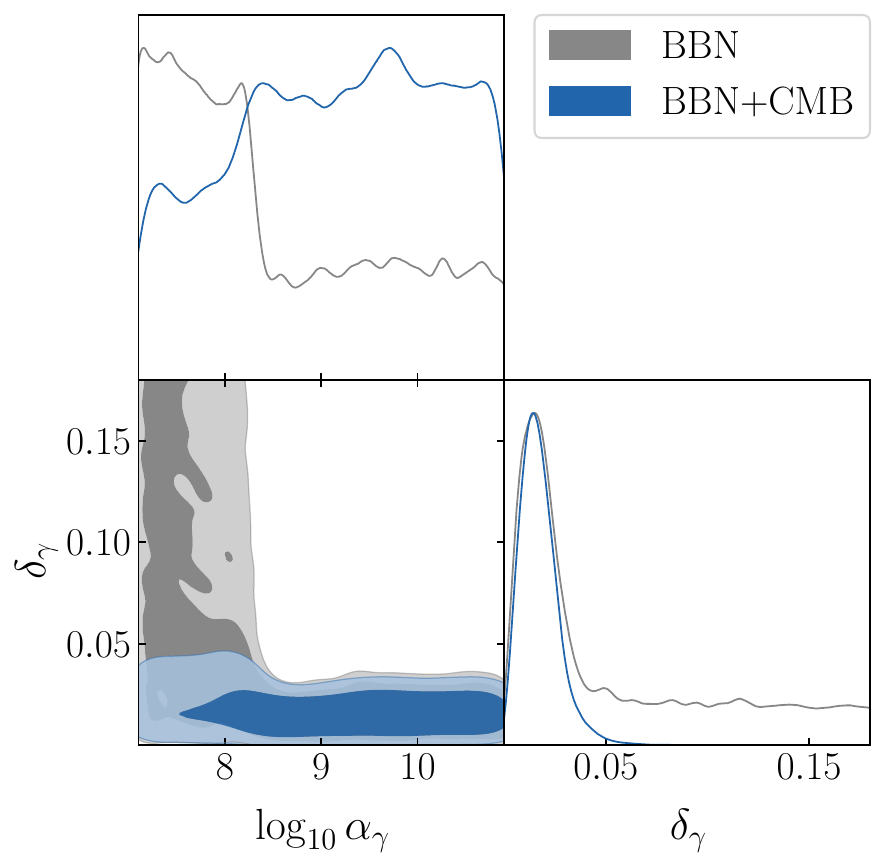} \caption{Same as Figure \ref{fig:Constraints_Tnu}, except for the scenario in which photon temperature is altered compared to its standard-model prediction, with parameters $\{\log_{10}\alpha_\gamma, \delta_\gamma \}$.}\label{fig:Constraints_T}
\end{figure}
%%%%%%%%%%%%

%%%%%%%%%%%%
\begin{figure*}
\includegraphics[width=0.46\textwidth]{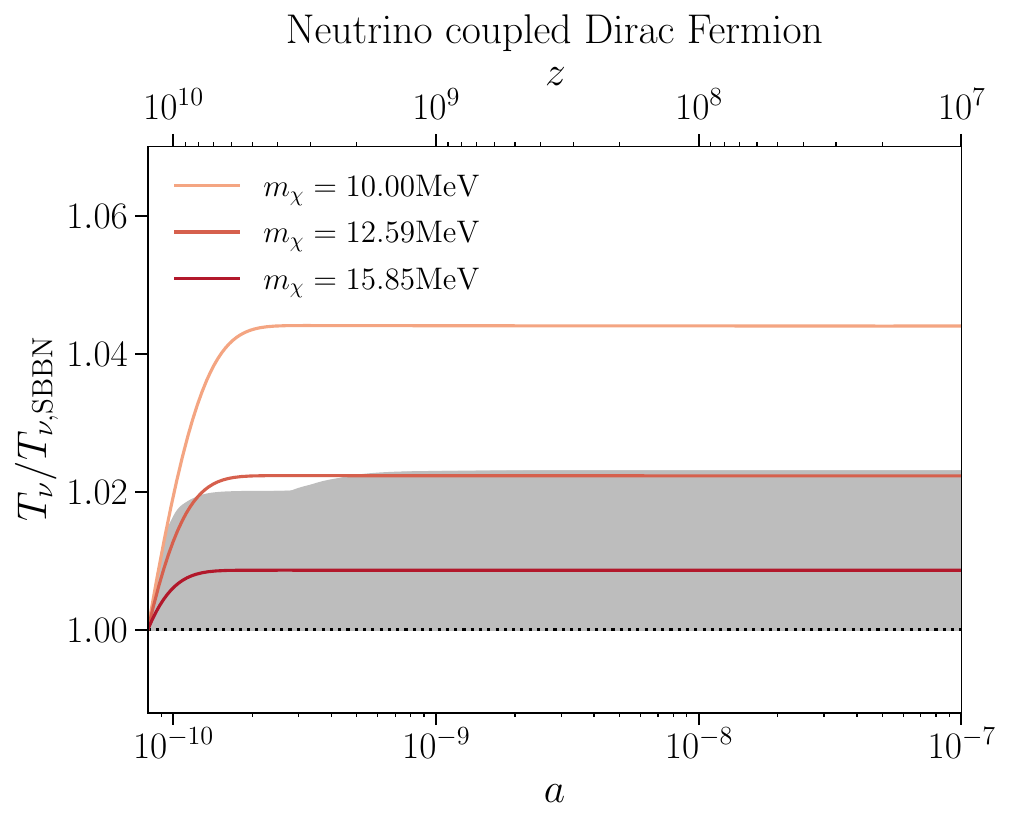}
\includegraphics[width=0.46\textwidth]{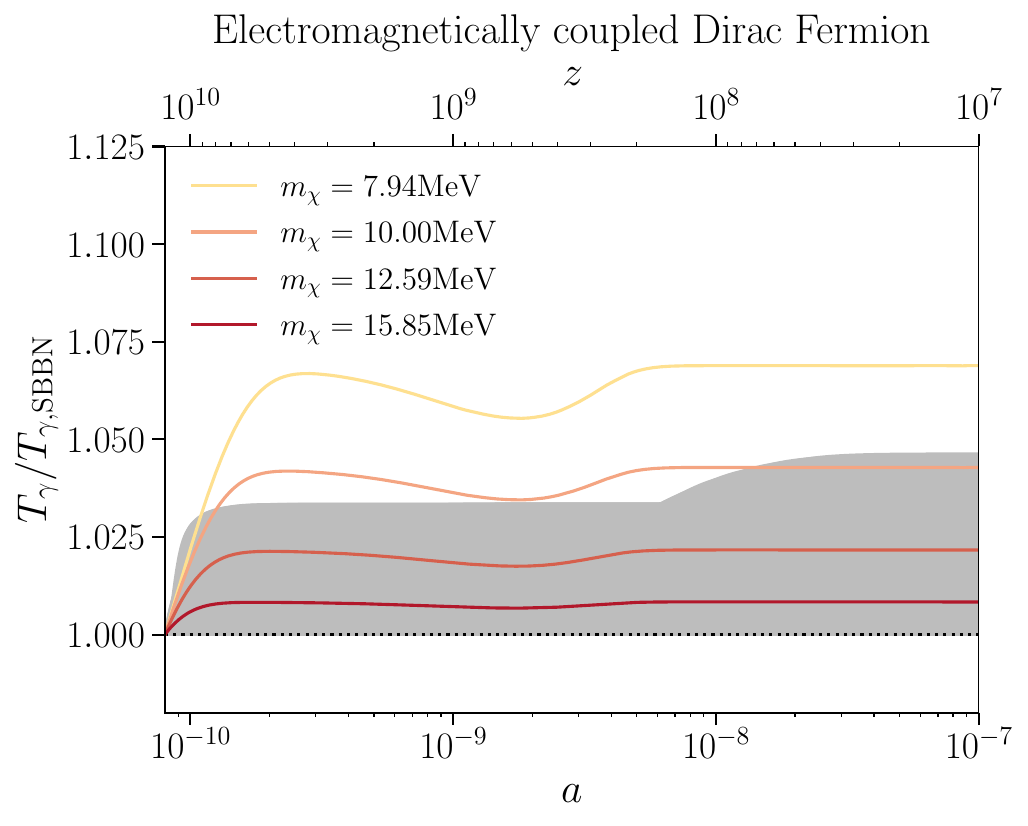}
\caption{The ratio of the temperature for a cosmology featuring light thermally coupled Dirac Fermion DM model, with a given particle mass $m_{\chi}$ shown in the legend, to the standard BBN scenario. The two panels correspond to electromagnetically coupled (right panel) and neutrino coupled (left panel) DM. The gray area shows the $95\%$ CL region for the radiation temperature, as inferred from the measurements of BBN + CMB. The DM masses that drive the ratio outside of this region are inconsistent with the data.}\label{fig:Mapping}
\end{figure*}
%%%%%%%%%%%%

In addition to the measurements of primordial abundances, the entropy dump processes affect the overall budget of radiation, quantified by $N_\mathrm{eff}$, defined by the ratio of neutrino to photon temperature, 
\begin{equation}
    N_\mathrm{eff}\equiv 3\left[\frac{11}{4}\left(\frac{T_\nu}{T_\gamma}\right)_0^3\right]^\frac{4}{3}\ ,
\end{equation}
where the subscript zero denotes the present time. %For example, particle annihilating into photons after neutrino decoupling can heat photons relative to the decoupled neutrinos, reducing the present-day ratio of neutrino-to-photon temperature, corresponding to a reduced value of $N_\mathrm{eff}$. Conversely, particle annihilating to neutrinos heats up neutrinos relative to photons, increasing $N_\mathrm{eff}$. 
In standard cosmology, with only the three SM neutrino species and SBBN, $N_\mathrm{eff} = 3.046$. In our empirical model where radiation temperature is altered, we have $N_\mathrm{eff} = 3.046/(1+\delta_\gamma)^4$ for the scenario where the photon temperature is altered, and $N_\mathrm{eff} = 3.046*(1+\delta_{\nu})^4$ for the case where the neutrino temperature is altered.

This time we do not limit ourselves to direct probes of the early universe only. In addition to the BBN yield measurements, we also consider the CMB measurement of $N_\text{eff}$ and $Y_\mathrm{p}$ from \textit{Planck} \cite{2020A&A...641A...6P}. We construct a chi-squared statistic $\chi^2_\text{tot} = \chi^2_\text{BBN}+\chi^2_\text{CMB}$ that depends on the model parameters (see details in Appendix~\ref{appendix.A}), and carry out the MCMC analysis to sample the posterior distribution for $\{\log_{10}\alpha_x, \beta_x, \gamma_x, \delta_x\}$. 

In Figures \ref{fig:Constraints_Tnu} and \ref{fig:Constraints_T}, we present the resulting constraints, where $\beta_x$ and $\gamma_x$ are marginalised over; we consider scenarios where neutrinos and photons are heated, respectively. The data cannot constrain $\beta_x$ and $\gamma_x$, which means the BBN yields and the CMB have no sensitivity to the speed of the annihilation process. From Figure \ref{fig:Constraints_Tnu}, we see that the BBN yields place an upper limit on the amount of the energy transferred to neutrinos $\delta_\nu$, and the bounds on $\delta_\nu$ mainly depend on the values of $\log_{10}\alpha_\nu$: the entropy dump process ends later as $\log_{10}\alpha_\nu$ goes to lower values. If this process happens after the end of BBN, corresponding to $\log_{10}\alpha_\nu \lesssim 8.2$, its impact on the neutrino temperature does not alter the primordial abundances, thus leaving $\delta_\nu$ unconstrained. From Figure \ref{fig:Constraints_T}, for the case of photon temperature, we again see that BBN observations can place a tight upper limit on $\delta_\gamma$ only in the range of $\log_{10}\alpha_\gamma \gtrsim 8.2$. In the same figures, we see that the CMB data plays a critical role in reducing the degeneracy between $\delta_x$ and $\alpha_x$, and we observe a significant improvement in the constraining power on $\delta_x$ when the CMB data is added. This is especially true for processes happening at later times $\log_{10}\alpha_x \lesssim 8.2$, which are sensitively probed by the measurement of $N_\mathrm{eff}$. 

Even though we used DM annihilation scenarios to check the flexibility of our empirical model in capturing modifications to the standard thermal history, the bounds we derive on $\{\log_{10}\alpha_x, \beta_x, \gamma_x, \delta_x \}$ do not depend on assumptions about the specific particle content of the early universe, and can be mapped to the constraints on any specific particle models of interest. As an example, in Figure \ref{fig:Mapping} we compare the temperature deviations in the light thermally coupled Dirac Fermion DM models (colored lines) to the deviations allowed by the model-independent BBN+CMB constraints at 95\% CL (gray regions). The allowed values of $m_{\chi}$ correspond to the parameter space where the theoretical curve is consistent with the gray shaded regions, the values of $m_{\chi}$ outside these regions are excluded by our analyses. Using this comparison, we get a bound for neutrino coupled species $m_{\chi}\gtrsim 12.6$ MeV, and $m_{\chi}\gtrsim 10.8$ MeV as the bound for electromagnetically coupled species, which are consistent with previous model-specific analyses \cite{2020JCAP...01..004S, 2022JCAP...07..002A}. 

%%%%%%%%%%%%%%%%%%%%%%%%%%%%
\section{Summary and Discussion}\label{sec:conclution}

We performed a model-independent reconstruction of the early-universe expansion and thermal histories, using measurments of the primordial element abundances, supplemented with the CMB measurements of the effective number of relativistic species and helium-4 abundance. For this purpose, we used two approaches. 

First, we adopted a non-parametric approach that allows for deviations from the standard-model values of $H(z)$ and $T(z)$, in a number of redshift bins. Using Fisher matrix analysis, we found that percent-level constraints on the expansion rate can be placed by the BBN yield data around the onset of nuclear interactions. For temperature evolution, the same data additionally place tight constraints around the epoch of neutrino decoupling. Guided by the results of the Fisher analysis, we then used the MCMC analysis to place bounds on CMB temperature evolution, as well as the expansion rate, in the early universe (at $z\gtrsim 10^7$), using light element abundances; our key results are shown in Figures~\ref{fig:CMB_Tem} and \ref{fig:expansion}. We study the primordial lithium problem and find that large deviations to the early expansion history around the onset of nuclear interactions might be a possible solution to the lithium problem.

Second, we used an empirical, physically-motivated model to capture deviations in the expansion history and radiation temperature in the early universe from the standard cosmological model. We performed an MCMC fit to the BBN yield data to find percent-level bounds on the matter energy density and 15\% bounds on the radiation energy density around BBN under the model-independent assumptions, using the measurements of primordial helium-4 and deuterium abundances. 
We further adopted a new empirical parameterization for the radiation temperature evolution, given in Eq.~(\ref{eq.paramH}), and based on parameters that regulate the timing and the speed of the entropy dump into a given radiation component, as well as the amount of the energy transferred (from annihilations, decays, or other processes). We have explicitly shown that this parameterization can flexibly capture a broad range of DM annihilation scenarios considered in the literature. We derived constraints on the free parameters of this empirical model using the BBN yield measurements of helium-4 and deuterium, as well as the measurements of $N_\mathrm{eff}$ and $Y_\mathrm{p}$ derived from CMB. We find that the element abundances alone can place an upper limit on the amount of the transferred energy, if the entropy dump occurs before the end of BBN. Adding the CMB data can significantly improve the constraints on the corresponding parameters, especially for annihilation processes happening at later times. However, we note that the two data sets probe very different times in cosmic history: while the BBN yields directly probe the early universe at an MeV energy scale, the CMB probes much later times. We show the key constraints in Figures \ref{fig:Constraints_Tnu} and \ref{fig:Constraints_T}.

We note that the reconstruction of the thermal and expansion histories presented in this study is model independent, and can thus conveniently map onto any constraints on any specific particle models of interest to cosmology. Within the landscape of new-physics models and in the era of the ever-increasing volume of cosmological data sets that can test them, our results aim to provide clarity on the most fundamental properties of the early universe, reconstructed with minimal assumptions about the unknown physics that can occur at these energy scales.

\section*{Acknowledgements}
RA and VG acknowledge the support from NASA through the Astrophysics Theory Program, Award Number 21-ATP21-0135. VG acknowledges the support from the National Science Foundation (NSF) under CAREER Grant No. PHY-2239205 and the support from the Cottrell Scholars Award from the Research Corporation for Science Advancement.

\clearpage

\appendix
\onecolumngrid

\section{Treatment of BBN and CMB Covariances}\label{appendix.A}
In this appendix, we detail how to obtain the BBN and CMB constraints outlined in the main text. To perform the fit on a model with some free parameters parametrized by $\boldsymbol{\theta}$, we factor the likelihood as $\mathcal{L} = \mathcal{L}_\mathrm{prob}\times \mathcal{L}_\mathrm{pri}$, where $\mathcal{L}_\mathrm{prob} \propto \mathrm{exp}\{-\frac{{\chi}^2}{2}\}$ is the probability likelihood, and $\mathcal{L}_\mathrm{pri}$ is the prior likelihood. 

For BBN data only, we compute the chi-squared statistic as 
\begin{equation}
    \chi^2_{\text{BBN}} = \left(\textbf{X} - \textbf{X}_{\rm{obs}}\right)^T \textbf{Cov}_{\mathrm{BBN}}^{-1} \left(\textbf{X} - \textbf{X}_{\mathrm{obs}}\right) ,
\end{equation}
where $\textbf{X}=(Y_\mathrm{p}, Y_\mathrm{D})$ is a vector of primordial element abundances, as a function of $\boldsymbol{\theta}$; $\textbf{X}_{\rm{obs}}$ is the vector of observed central values of the corresponding elements~\cite{2020PTEP.2020h3C01P}, giving $\textbf{X}_{\rm{obs}}=(0.245, 2.547)$; and $\textbf{Cov}_{\mathrm{BBN}}$ is the covariance matrix for the element abundances we are considering. 

There are two contributions to $\textbf{Cov}_{\mathrm{BBN}}$: one is from the observed uncertainties, which can be straightforwardly obtained from the BBN measurements in Ref.~\cite{2020PTEP.2020h3C01P}, giving $\sigma_{Y_\mathrm{p}}^\mathrm{obs}=0.003$ and $\sigma_{Y_\mathrm{D}}^\mathrm{obs}=0.025$; the other one is from the theoretical uncertainties in the computation of elemental abundances, which arise from uncertainties in the neutron lifetime and various nuclear reaction rates. The theoretical uncertainties can be estimated by BBN code--\texttt{AlterBBN}, giving $\sigma_{Y_\mathrm{p}}^\mathrm{th}=3.2\times10^{-4}$, $\sigma_{Y_\mathrm{p}}^\mathrm{th}=0.038$ and $\tau^\mathrm{th}_{Y_\mathrm{p}, Y_\mathrm{D}} = -0.012$, where $\tau^\mathrm{th}_{Y_\mathrm{p}, Y_\mathrm{D}}$ is the covariance between $Y_\mathrm{p}$ and $Y_\mathrm{D}$. We can add the observed and theoretical uncertainties in quadrature to obtain the full covariance matrix
\begin{alignat}{1}
     \textbf{Cov}_{\mathrm{BBN}} = \begin{bmatrix}
         (\sigma_{Y_\mathrm{p}}^\mathrm{th})^2 + (\sigma_{Y_\mathrm{p}}^\mathrm{obs})^2 & \tau^\mathrm{th}_{Y_\mathrm{p}, Y_\mathrm{D}} \\
         \tau^\mathrm{th}_{Y_\mathrm{p}, Y_\mathrm{D}} & (\sigma_{Y_\mathrm{D}}^\mathrm{th})^2 + (\sigma_{Y_\mathrm{D}}^\mathrm{obs})^2 
     \end{bmatrix} \, .\label{eq:BBN_cov_2}
\end{alignat}
This covariance can be extended to a 4-dimensional matrix if we include the measurements of the other two element abundances ${}^3\!He/H$ and ${}^7\!Li/H$, giving 
\begin{alignat}{1}
     \textbf{Cov}_{\mathrm{BBN}} = \begin{bmatrix}
         (\sigma_{Y_\mathrm{p}}^\mathrm{th})^2 + (\sigma_{Y_\mathrm{p}}^\mathrm{obs})^2 & \tau^\mathrm{th}_{Y_\mathrm{p}, Y_\mathrm{D}} & \tau^\mathrm{th}_{Y_\mathrm{p}, {}^3\!He/H} & \tau^\mathrm{th}_{Y_\mathrm{p}, {}^7\!Li/H} \\
         \tau^\mathrm{th}_{Y_\mathrm{p}, Y_\mathrm{D}} & (\sigma_{Y_\mathrm{D}}^\mathrm{th})^2 + (\sigma_{Y_\mathrm{D}}^\mathrm{obs})^2 & \tau^\mathrm{th}_{Y_\mathrm{D}, {}^3\!He/H} & \tau^\mathrm{th}_{Y_\mathrm{D}, {}^7\!Li/H} \\
         \tau^\mathrm{th}_{Y_\mathrm{p}, {}^3\!He/H} &
         \tau^\mathrm{th}_{Y_\mathrm{D}, {}^3\!He/H} & (\sigma_{{}^3\!He/H}^\mathrm{th})^2 + (\sigma_{{}^3\!He/H}^\mathrm{obs})^2 & \tau^\mathrm{th}_{{}^3\!He/H, {}^7\!Li/H} \\
         \tau^\mathrm{th}_{Y_\mathrm{p},{}^7\!Li/H} &
         \tau^\mathrm{th}_{Y_\mathrm{D},{}^7\!Li/H} &
         \tau^\mathrm{th}_{{}^3\!He/H,{}^7\!Li/H} &(\sigma_{{}^7\!Li/H}^\mathrm{th})^2 + (\sigma_{{}^7\!Li/H}^\mathrm{obs})^2 
     \end{bmatrix} \, .\label{eq:BBN_cov_4}
\end{alignat}
where $\sigma_{{}^3\!He/H}^\mathrm{obs}=2\times10^{-6}$, $\sigma_{{}^7\!Li/H}^\mathrm{obs}=3\times10^{-11}$~\cite{2020PTEP.2020h3C01P}, and the theoretical uncertainties can be estimated by BBN code. 

To assess the consistency of a given model with the measurements of BBN and CMB, we construct a chi-squared statistic which is a sum of two separate chi-squared statistics for each observable:
\begin{equation}
    \chi^2_{\mathrm{tot}} = \chi^2_{\mathrm{BBN}} + \chi^2_{\mathrm{CMB}}\ . 
\end{equation}
The CMB contribution is 
\begin{equation}
    \chi^2_{\mathrm{CMB}} = \left(\textbf{Y} - \textbf{Y}_{\rm{obs}}\right)^{\mathrm{T}} \textbf{Cov}_\mathrm{CMB}^{-1} \left(\textbf{Y} - \textbf{Y}_{\mathrm{obs}}\right) , 
\end{equation}
where $\textbf{Y}=(N_\mathrm{eff}, Y_\mathrm{p})$ as a function of $\boldsymbol{\theta}$; $\textbf{Y}_{\rm{obs}}$ is the central value of these two parameters derived from the CMB measurements; and $\textbf{Cov}_\mathrm{CMB}$ is the covariance matrix. We use the low-$\ell$ and high-$\ell$ multifrequency power spectra TT, TE, and EE from \textit{Planck} PR3 (2018)~\cite{2020A&A...641A...6P}, and carry out a MCMC analysis within the \texttt{Cobaya} sampling framework~\cite{2021JCAP...05..057T,2019ascl.soft10019T} to determine the values of $N_\mathrm{eff}$ and $Y_\mathrm{p}$ that are consistent with the CMB measurements. For this dataset, we have

\begin{alignat}{1}
    \textbf{Y}_{\rm{obs}} = (N^{\rm{obs}}_\mathrm{eff}, Y^{\rm{obs}}_\mathrm{p}) = (2.85, 0.2473) \,, \qquad \textbf{Cov}_\mathrm{CMB} = \begin{bmatrix}
         6.8 \times 10^{-2}  & -2.9 \times 10^{-3}\\
         -2.9 \times 10^{-3} & 2.3 \times 10^{-4}
    \end{bmatrix} \,.
\end{alignat}

\clearpage
\section{Full probability distributions}\label{appendix.B}

\begin{figure*}[!h]
\includegraphics[width=0.98\textwidth]{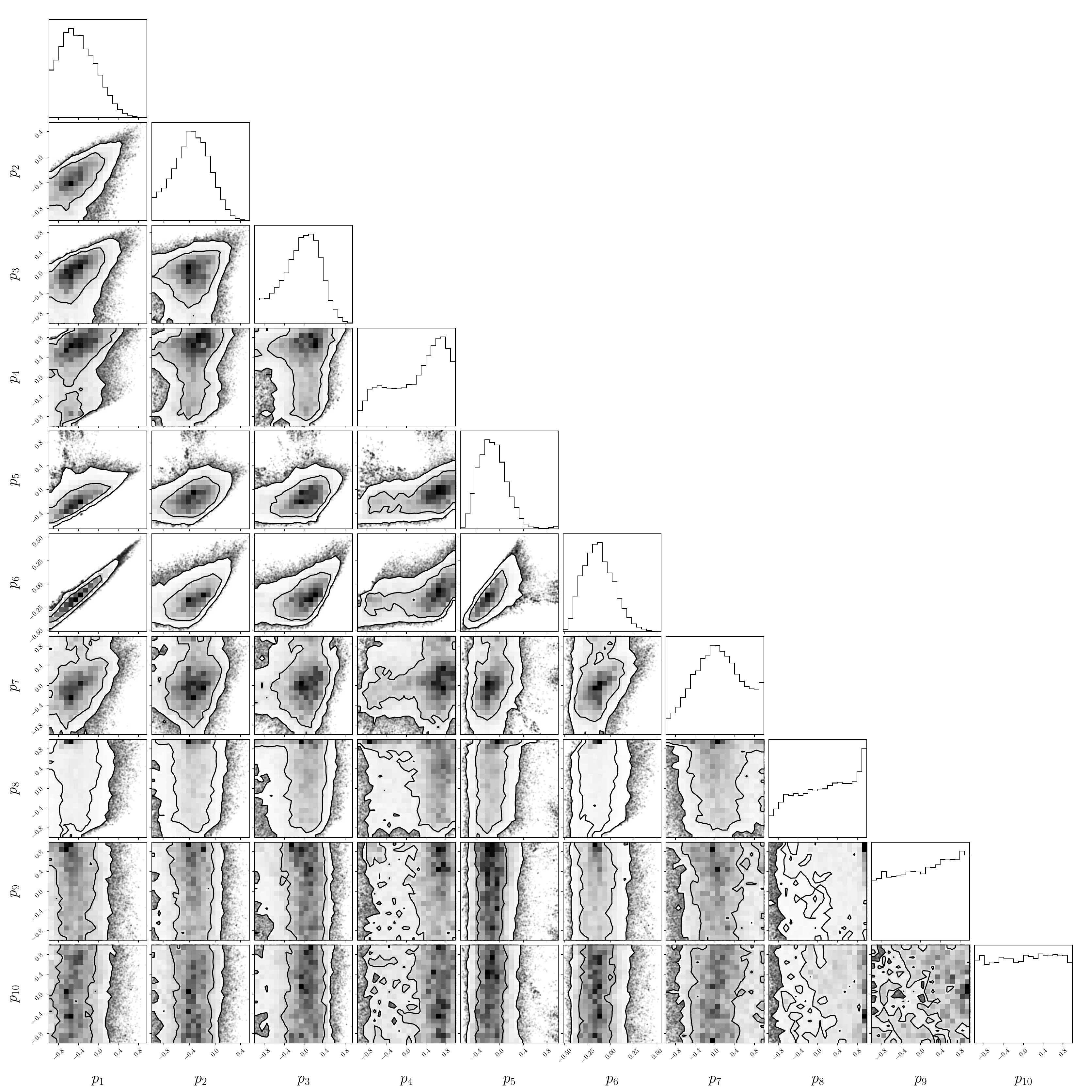}
\caption{The probability distribution for the 10 bin parameters $\{p_i\}$ in our model-independent analysis for CMB temperature. We show the 68\% and 95\% C.~L.~contours obtained from BBN measurements of $Y_\mathrm{p}$ and $Y_\mathrm{D}$. The one-dimensional marginalized posteriors are shown at the top of each column.}\label{fig:fullPD_T}
\end{figure*}

We show the full marginalized posterior distributions for the 10 bin parameters $\{p_i\}$ in our model-independent analysis for CMB temperature in Fig.~\ref{fig:fullPD_T}. 

%%%%%%%%%%%%
%%%%%%%%%%%%

\clearpage
\twocolumngrid

\bibliographystyle{IEEEtran}
\bibliography{wimp}

\end{document}